\documentclass[12pt,a4paper,final]{iopart}

\usepackage{iopams}  
\usepackage{graphicx}
\usepackage{amssymb}
\usepackage{cite}

\usepackage[breaklinks=true,colorlinks=true,linkcolor=blue,urlcolor=blue,citecolor=blue]{hyperref}

\DeclareGraphicsExtensions{.pdf,.jpg, .png}
\usepackage{epstopdf}
\usepackage{enumitem}
\usepackage{subcaption}
\usepackage[font=footnotesize]{caption, subcaption}
\usepackage{bm}

\begin{document}
\title[]{On the Changes in Optical Interferometry Induced by The Relativistic Motion of An Optical Medium \let\thefootnote\relax\footnote{Accepted for publication in \textit{European Journal of Physics}.\\
Accepted: 12 August, 2016. Published: 22 September 2016.  }
 }
\author{Supantho Rakshit}
\address{Notre Dame College\\ Motijheel Circular Road, Arambagh, Motijheel, Dhaka-1000, Bangladesh }
\eads{\mailto{sukantoraxit@gmail.com}}
\begin{abstract}
We investigate the effects of relativistic movement of optical medium on the conditions of constructive and destructive interference, reflection and transmission pattern, and performance of spectroscopes. First, we consider the case of two beams reflected from a relativistically moving thin film and derive the conditions of constructive and destructive interference. Then we broaden the idea to multiple beam interference and formulate a new modified equation of reflection and transmission pattern of light from a relativistically moving plane parallel dielectric film. Finally, we determine the new effective resolving power of a Fabry Perot spectroscope, which has a moving dielectric medium in its etalon.  Here, we consider the case where the optical mediums move parallel to the plane of incidence of light. Throughout this paper, we use basic Lorentz transformation of space and time co-ordinates and electromagnetic fields to conduct these investigations. This paper can strongly motivate undergraduate physics students to associate the concepts of special relativity and optical interference.
\end{abstract}

\vspace{2pc}


\section*{Introduction}
\label{intro}

In most undergraduate physics courses, the topics of special relativity and optical interference are taught in a complete separate manner. The overall schemes of these two topics might seem somewhat distant to an undergraduate student. For example, topics like Lorentz transformation and thin film interference do seem like they are completely unconnected. So, in most of the times, both the process of teaching and learning of these two very important and interesting topics remain completely isolated.\\
But, the key ideas of these two topics can be merged together, if we switch to a frame where one or more of the optical media are moving (hypothetically, in relativistic regime). In fact, the association of special relativity and optics is nothing new. The problem of reflection and transmission of electromagnetic wave from a moving dielectric medium is one of the fundamental problem of electrodynamics of moving media. It has been explored and explained ingreat details throughout the years. Major works include Pauli \cite{Pauli} and Sommerfeld \cite{sommerfeld1959vorlesungen}, on the frequency shift of the reflected wave by a moving mirror. Later Tai \cite{tai1965ursi} and Yeh \cite{yeh1965reflection} provided the solutions for reflection and transmission co-efficients of plane electromagnetic waves by a dielectric half space in vacuum, moving parallel and perpendicular to the interface respectively. Then, the problem has been discussed in further details, such as: reflection and refraction of electromagnetic waves by a moving dielectric medium moving in an arbitrary direction parallel to the plane of incidence \cite{pyati1967reflection}, by a dielectric half space moving perpendicular to the plane of incidence \cite{shiozawa1967reflection}, for plane waves reflected from an arbitrarily moving medium \cite{Huang}. Other works include finding the Brewster angle for a dielectric medium moving in an arbitrary direction \cite{brewster}, the total reflection at the interface between two relatively moving dielectric mediums \cite{shiozawa1967total}, Snell's law for the Poynting vector in a semi infinite dielectric medium moving perpendicular  to the surface etc \cite{snell} (to name only a few).\\
Some extended the investigation to many other aspects, such as the scattering of plane waves at a plane interface separating two moving media \cite{scattering}; the reflection and transmission of electromagntic wave by a moving inhomogeneous medium \cite{inhomogenous}, by a ferrite surface \cite{ferrite}, and a moving plasma medium \cite{plasma} etc. Ref. \cite{1999optics} also discusses some very interesting optical properties of non-uniformly moving media.\\
In this paper, the most basic generalizations of those ideas are accumulated and integrated to optical interferometry by conducting some simple analysis. These would help gain some insights as to the modifications brought about in various aspects of optical interference by the relativistic motion. Here we associate the basic concepts of special relativity to a moving optical medium to investigate the followings:\\
(i) the conditions of constructive and destructive interference in the case of a moving thin film, and\\
(ii) the reflection and transmission pattern of light from a moving dielectric plane parallel film and how it compares with the non-moving situation.\\
We will limit our concerns for motion parallel to the plane of incidence of light.
After the aforementioned topics have been analyzed, we then use the results to address a more practical concern:\\
How does a moving dielectric slab in a Fabry Perot (FP) etalon affects the spectroscope? Is there any change in the reflection or transmission pattern? Does the movement of the medium inside the etalon somehow affects the performance of the spectroscope by changing its resolving power?\\
We discuss FP spectroscope because it's functions are built on the concepts of multiple beam interference, which was the issue of interest prior to this, and also because it is one of the most widely taught interferometers in undergraduate level alongside Michaelson-Morley interferometer, N-slit interferometer, Fresnel biprism etc. It is also amongst the most used in the field of optical research \cite{FPbook}. \\
All the calculations throughout the paper are carried out using the most basic physics and mathematics taught at undergraduate level, therefore making it extremely useful for any young undergraduate student to merge the concepts of relativity and optical interference. This paper may also inspire the student to investigate other optical properties of medium that are affected by the relativistic movement.  \\
We have divided the overall discussions into three sections.
In section \ref{sec1}, we explore the two beam interference conditions for a moving thin film. Then using the results obtained in section \ref{sec1}, we expand the idea to multiple beam interference in section \ref{sec2}. Here, we derive an expression for the new reflection and transmission pattern. Then finally in section \ref{sec3}, we discuss the FP spectroscope, which has a moving medium in its etalon. We derive the new modified conditions of interference and also the new resolving power. In each and every step, we will check the the validity of the derived expressions by examining whether they reduce back to their familiar forms in non-moving situations. Thus we will complete an overall inspection on the effects of relativistic movement of optical medium in interference.\\
\section{Interference of Two Light Beams Due to Reflection From a Moving Thin Film}
\label{sec1}
Consider a thin film of thickness $d$ moving at a speed $v$ with respect to the lab frame $ S$. A light source in air is located at an angle $i$ to the vertical. It is stationary in $S$. It impinges light of frequency $\nu_{0}$, measured in lab frame $S$, on the film as shown in Fig. 1. All the points on the film will act as sources of secondary wavelets, according to the Huygen-Fresnel principle \cite{born}.

\begin{figure}[h]
\label{fig1}
\includegraphics[scale=.20]{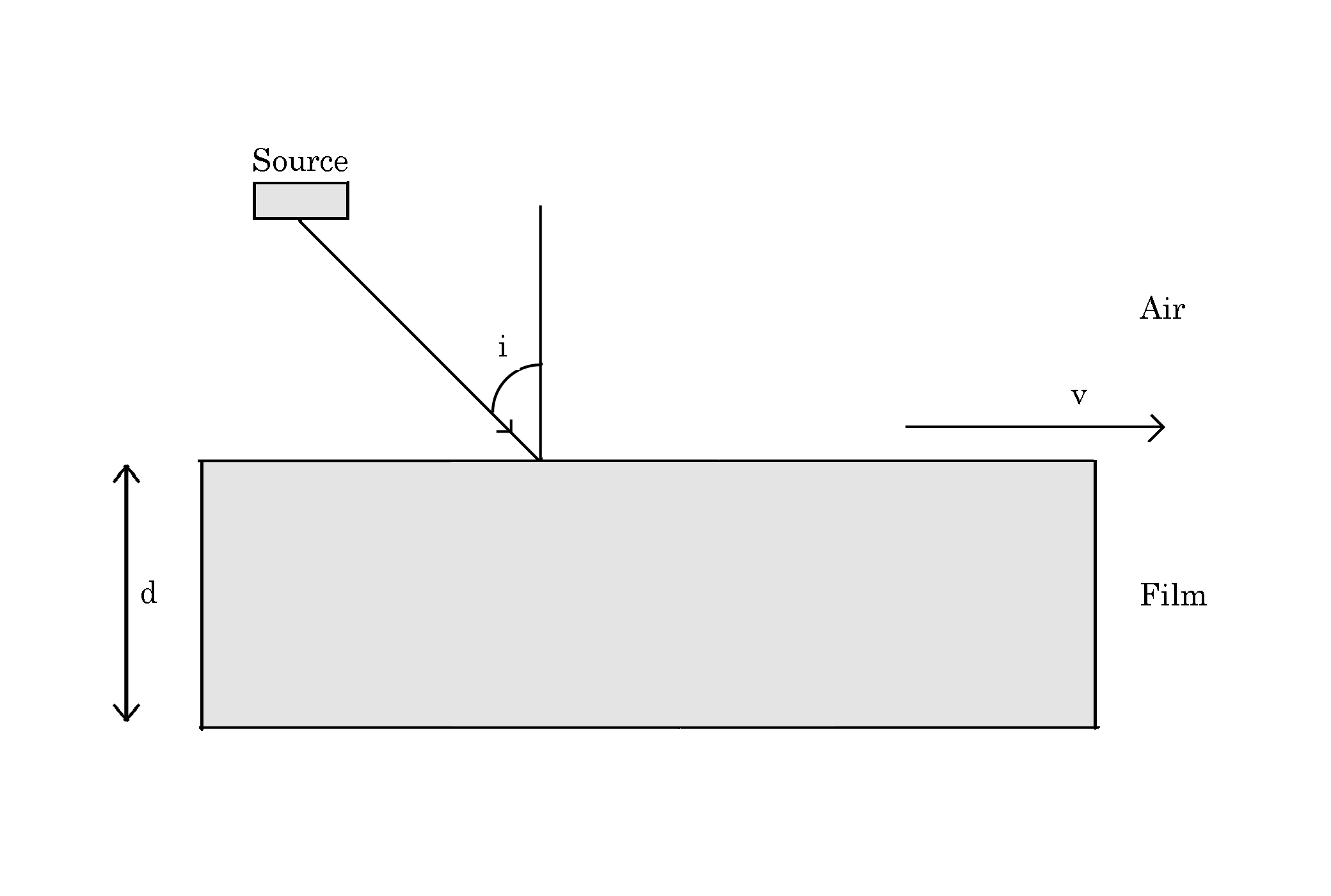}
\centering

\caption{A thin film of thickness $d$ is seen moving rightwards at speed $ v $ in lab frame $S$. The light ray from a source situated in air impinges on the film at an angle $i$ to the vertical (as measured in $ S $)}
\end{figure}
\newpage

In the frame of the film $S'$, the light source is seen moving backwards at speed $v$. So, in $S'$ the incoming light is Doppler shifted. So the relativistic Doppler shifted frequency \cite{kleppner} is \\
\begin{equation}
\label{eq1}
\nu' =\nu_0\frac{\sqrt{1-\beta^2}}{1+\beta\ \sin \ i'}      .
\end{equation}
Here $i'$ is the angle that the incoming light wave makes with the vertical as seen in $S'$ and $\beta=v/c$.\\

Using relativistic velocity transformation, it follows that\\
\begin{equation}
\label{eq2}
 \tan i'=  \gamma  \frac{\left(\sin i-\beta\\ \right)}{\cos i}   .
 \end{equation}
Where  $\gamma=1/\sqrt{1-\beta ^2} $ is the Lorentz factor. If $n'$ is the refractive index of the film measured in $S'$, using Snell's law in $S'$, we have
\begin{equation} 
\label{eq3} 
\sin i'=n'\ \sin r'\end{equation} or,
\begin{equation}
\label{eq4}
\sin r'=\frac{\frac{\gamma}{n'}\left(\sin i-\beta\right)}{\sqrt{\cos ^2i+\gamma^2\left(\sin i-\beta\ \ \right)^2}}    .
\end{equation}

Keeping in mind that the medium of the film is dispersive, the refractive index $n'$ used in Eq. (\ref{eq3}) is the refractive index that corresponds to frequency $ \nu' $, the frequency as measured by $S'$. The relation between $ n' $ and $ \lambda^{'}$ ($\lambda'$ is the corresponding wavelength as measured in frame $S'$ which equals $c/\nu'$), can be easily obtained using the empirical Cauchy relation, which  is $ n'^{2}= A+ B/{\lambda'}^{2}$, where $ A $ and $ B $ are two constants \cite{white}.\\
Using straightforward inverse Lorentz transformation to find the refraction angle $r$ in $S$, we have
\begin{equation}
\label{eq5}
\tan \ r=\frac{\ \gamma^2\left(\sin \ i-\beta\right)+\gamma \beta n'^{2} c  \sqrt{\cos ^2i+\gamma^2\left(\sin \ i- \beta \right)^2}}{n'\ \sqrt{\cos ^2i+\gamma^2\left(\sin \ i-\beta \right)^2\left(1-\frac{1}{n'^2}\right)^{ }}}.
\end{equation} 
The effective refractive index of the moving film $ n_{eff} $, measured in $S$ can be defined as\begin{normalsize}
$ n_{eff}= \sin i/ \sin r $.
\end{normalsize}As the angle $r$ in Eq. ($5$) is a function of $ \gamma $ and $i$, $ n_{eff} $ can be writtten as a function of $r$ and $\gamma $. Therefore, the speed of light measured by $S$ within the film will depend on the direction of the light wave, making the medium of the film to look anisotropic, even if the medium is isotropic when at rest with respect to $S$. For a detailed analysis of this anisotropy, see \cite{anisotropy}. \\

The overall phenomenon observed by $S'$ is shown in Fig. 2.\\
\begin{figure}[h]
\label{fig2}
\includegraphics[scale=.21]{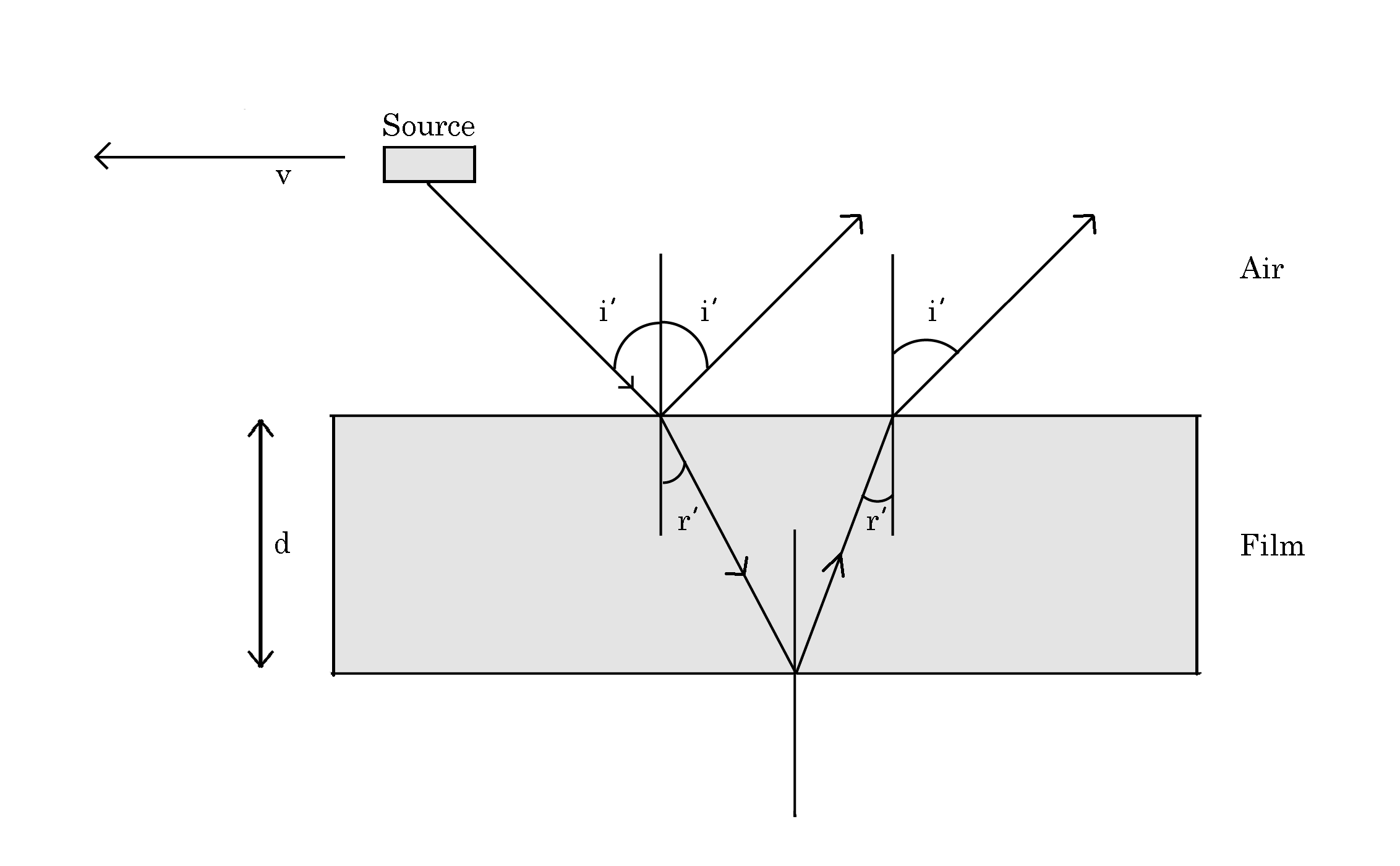}
\centering

\caption{The light source is seen to move leftwards with speed $v$ in frame $S'$. The light ray is incident at an angle $i'$ to the vertical in $S'$. The ray splits into two. One of them immediately reflects back to air at the same angle $i'$. Another refracts through the film at an angle $r'$, reaches the bottom of the film, reflects from there and finally re-transmits to air at angle $i'$ to the vertical.}
\end{figure}

We will conduct some simple relativistic treatments now. In frame $S$ and $S'$, when measured time $ t=t'=0 $, the light ray falls on the moving film. Some of the light is reflected back to air and rest is transmitted through the medium (we neglect any absorption loss). And, after a time interval $ \Delta t' $ in $S'$, the transmitted light ray again re-transmits back to air after getting reflected from the lower part of the film.  
So from Fig. 2, it follows that 
\begin{equation}
\label{eq6}
\Delta t'= \frac{2n'd}{c\cos r'}   
\end{equation}
and
\begin{equation}\label{eq7} 
\Delta x'=2d\tan r' .
\end{equation}

\noindent
In $S$, $\Delta x=\left(\Delta x'+\frac{\beta}{c} \Delta t'\right)\gamma =$ \begin{Large}
$\frac{2d\gamma\left(\sin r'+n'\beta\right)}{\cos r'}$
\end{Large}, and 

\begin{equation}
\label{eq8}
\Delta t=\left(\Delta t'+\frac{\beta\Delta x'}{c}\right)\gamma= \frac{2d\gamma}{c}\frac{\left(n'+\beta \sin r'\right)}{\cos r'} .
\end{equation}
 
\begin{figure}[h]
\includegraphics[scale=.2]{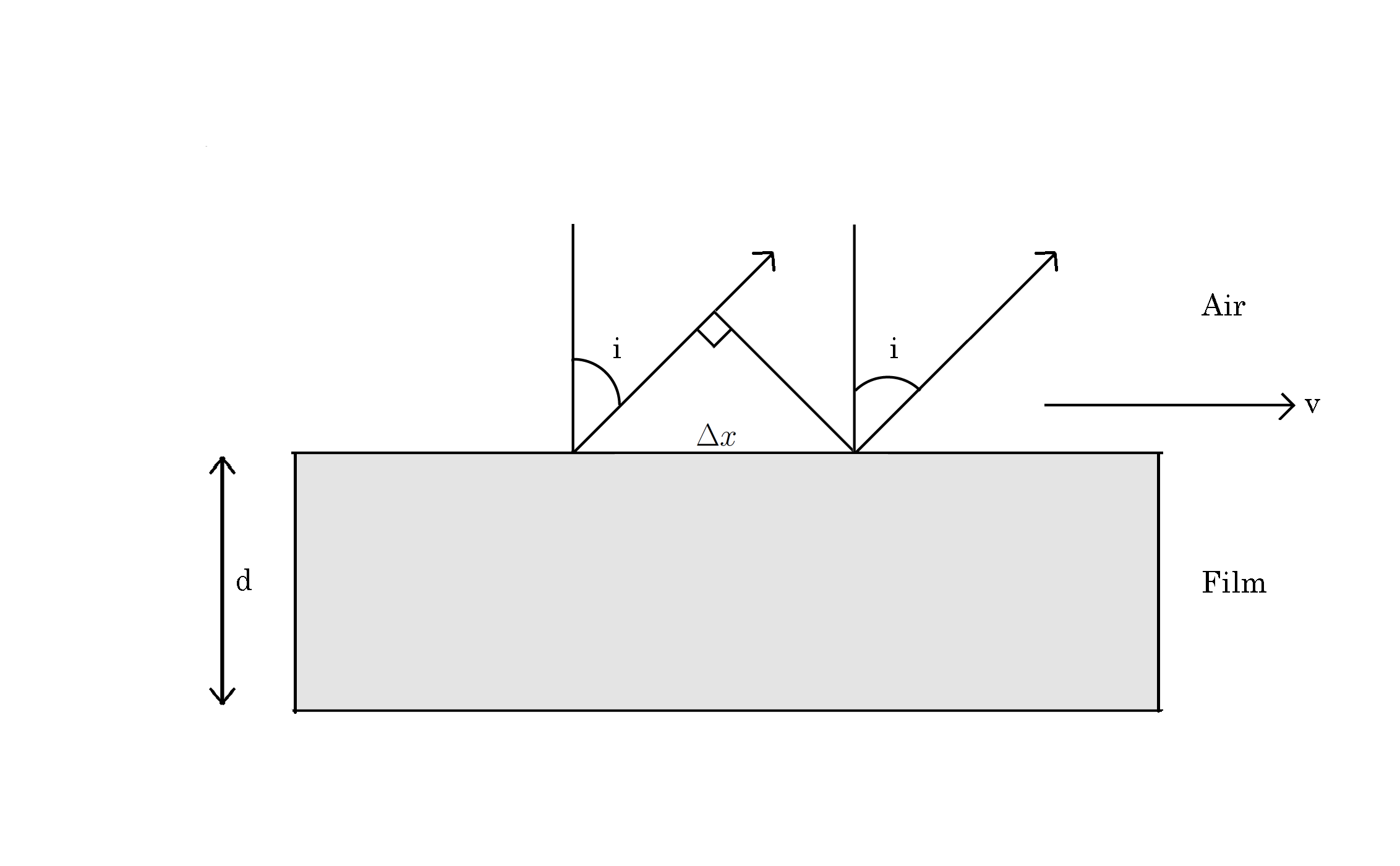}
\label{fig3}
\centering

\caption{The re-transmitted light rays as seen by frame $S$. The rays are at an angle $i$ to the vertical and separated a horizontal distance $\Delta x$ in that frame.}
\end{figure}

In $S'$, the incident light reflects at an angle $i'$, which transforms back to angle $i$  in $S$, as shown in Fig. 3. This result can also be derived from Einstein's formula of reflection from a moving mirror, which is $\sin i-\sin r=\beta \sin \phi \sin \left(i+r\right)$ \cite{apho}, where $ \phi$ is the angle the mirror makes with the direction of the relative velocity. In this case $\phi$ is zero, which leads to $i=r$. This law can be derived using the Huygens-Fresnel principle, as well as Lorentz transformations \cite{Gjurchinovski1}.

So the path difference between the two beam in $S$ is\\
\begin{center}
$\Delta=c\Delta t-\Delta x \sin i$   .
\end{center}
After substituting the values of $\Delta t$ and $\Delta x$, we find\\
\begin{equation}
\label{eq9}
\Delta = \frac{2d\gamma}{\cos r'}\left\{\left(n'+\beta\sin r'\right)-\sin i\ \left(\sin r'+n'\beta\right)\right\}        .
\end{equation}

Here, if $ \beta =0$, then $\Delta$ reduces to the simple cosine law, that is $\Delta =2nd \cos r $ \cite{white}.\\

There is another issue of interest here. The moving Huygens sources of the film generate light of frequency $\nu'$ in their own frame $ S' $. But the detector (situated in the lab frame) will see the sources moving rightwards at speed $v$, so the received light will again be Doppler shifted. The frequency of light received by the detector is
\begin{center}
\begin{equation}
\label{eq10}
\nu_r =\nu'\frac{\sqrt{1-\beta^2}}{1-\beta\ \sin \ i}
      =\nu_0\frac{\left(1-\beta^2\right)}{\left(1+\beta\sin i'\right)\left(1-\beta\sin i\right)}          .
\end{equation}
\end{center}
So, the corresponding wavelength is
\begin{center}
\begin{equation}
\label{eq11}
\lambda_r =\frac{c}{\nu_r}
      =\lambda_0\frac{\left(1+\beta\sin i'\right)\left(1-\beta\sin i\right)}{1-\beta^2}          
      \end{equation}
\end{center}
where $\lambda_{0}= c/\nu_{0}$, the wavelength of the light impinging on the moving film, as measured in $S$. Here, if $ \beta =0$, then $\lambda_r =\lambda_0 $, as expected.\\\\

So, the condition for constructive interference is
\begin{center}
$\Delta= (m+\frac{1}{2})\lambda_r$ where m = 0, 1, 2, 3... etc\end{center}

or,

\begin{equation}
\label{eq12}
\frac{2d\gamma}{\cos \ r'}\frac{\left\{\left(n'+\gamma\ \sin \ r'\right)-\sin \ i\ \left(\sin \ r'+n'\gamma\right)\right\}\left(1-\beta^{^2}\right)}{\left(1+\beta\sin \ i'\right)\left(1-\beta\ \sin \ i\right)\ }= \left(m+\frac{1}{2}\right)\lambda_{0}.
 \end{equation}
 
\vspace{2mm}
 
And the condition for destructive interference is

\begin{equation}
\label{eq13}
\frac{2d\gamma}{\cos \ r'}\frac{\left\{\left(n'+\gamma\ \sin \ r'\right)-\sin \ i\ \left(\sin \ r'+n'\gamma\right)\right\}\left(1-\beta^{^2}\right)}{\left(1+\beta\sin \ i'\right)\left(1-\beta\ \sin \ i\right)\ }= m\lambda_{0}.
 \end{equation}

\section{Reflection and Transmission Pattern of Light From a Moving Plane Parallel Dielectric Film}
\label{sec2}
In the previous section, all the calculations were only for two beams. Actually, there were many more beams emanating from the film other than the considered ones. All those other beams went through multiple number of reflections within the boundaries of the film. But we disregarded those beams assuming low reflectivity of the film. The reason is, as the other beams underwent too many reflections, it led to an almost vanishing amplitude of the beams. So, as the amplitudes of those beams were really small, they contributed extremely small to the detected intensity. For this reason, only the first two beams were taken into account as they both were reflected only once. The first and second beam only got reflected from the upper and lower boundary between the film and air respectively. Hence, they are the ones that contribute most dominantly to interference. However, if reflectivity is high, then contributions other beams will have to be taken into account too. This main goal of this section is to expand the ideas of section \ref{sec1} to multiple beam interference. At first, we will derive the effective reflectivity of a moving plane parallel dielectric film. Then aggregate all the contributions by phasor addition. By the end of this section, we will have an expression for the new reflection and transmission pattern for a moving dielectric plane parallel film, which is moving parallel to the plane of incidence of light. We will use plane polarized light, with electric field vector $\textbf{E} $ perpendicular to the plane of incidence to simplify calculations.\\

So, there is light wave impinging on a plane parallel  dielectric film of thickness $d$ moving rightwards with speed $v$ with respect to lab frame $S$. The $\textbf{E}$ field in $S$ is \textbf{s}-polarized. Hence,
  ${{\textbf{E}}_{\parallel}}=0$ and $ \left|{{\textbf{E}}_{\perp}}\right|=\left|\textbf{E}\right|$.\\

   And the components of the associated magnetic field $ \textbf{B} $ are
${{\textbf{B}}_{y}}=\textbf{B} \sin i$ and \\${{\textbf{B}}_{x}}=\textbf{B} \cos i$.  And also,  $\left|{{\textbf{E}}_{\perp}}\right|=\left|\textbf{B}\right|c$. See Fig. 4.\\
\begin{figure}[h]

\includegraphics[scale=.07]{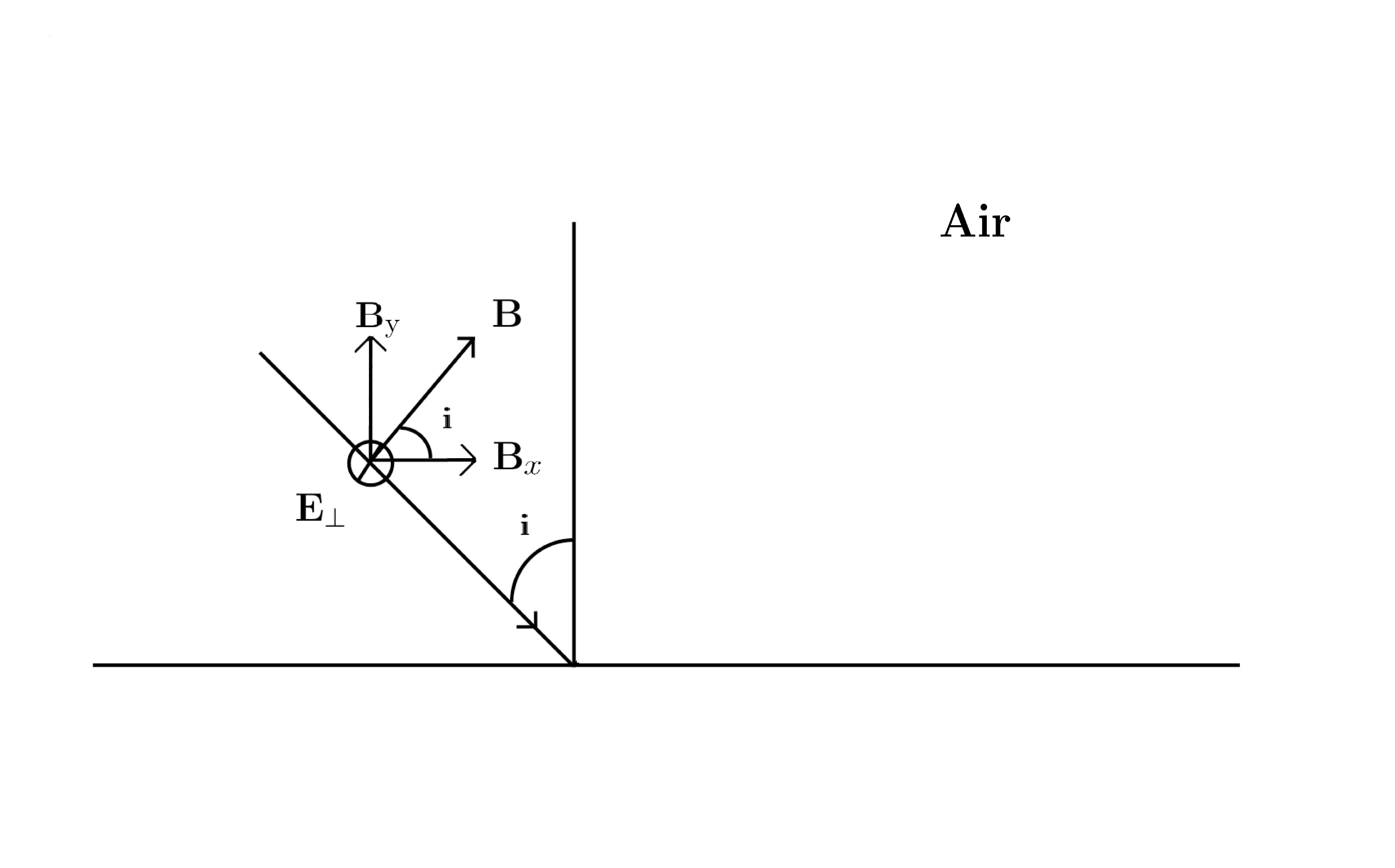}
\label{fig4}
\centering
\caption{The \textbf{E} and \textbf{B} vectors as seen in lab frame $S$. The \textbf{s}-polarized $\textbf{E} $ vector is perpendicular to the plane of incidence and points into the page. The associated $\textbf{B}$ is at an angle $i$ to the horizontal.}

\end{figure}

 The amplitude of \textbf{E} field observed by $S'$ is, by Lorentz transformation
	 \begin{equation}
	 \label{eq14}
	  E'_\perp  =\gamma\left(E_\perp-\beta cB_y\right).
\end{equation}	 
The minus sign arises from the fact that the direction of vector product of velocity and \boldsymbol{$B_y$} is in the opposite direction of \boldsymbol{$ E_\perp $}.\\
So now \begin{equation}
\label{eq15}
E'_\perp=\gamma\left(E_\perp-\beta Bc \sin i\right)
	=  \gamma E_\perp\left(1-\beta \sin i\right).
	\end{equation}
And the amplitude of the perpendicular component of \textbf{B} field, $B'_y$ observed by $S'$ is
\begin{equation*} B'_y=\frac{\gamma\left(c B_{y}-\beta E_\perp\right)}{c}
	= \frac{\gamma E_\perp\left(\sin i -\beta\right)}{c}
	\end{equation*}
or,
	\begin{equation}
	\label{eq16}
	  B'_{y}=\frac{\gamma E_\perp }{c}\left(\sin i\ -\beta\right) .
	\end{equation}

Now, we define certain quantities: $ r'$ and $ t'$ are the reflectivity and transmitivity. And $r'_{i}$ and $ t'_{i}$ are the `inverse` reflectivity and transmitivity, of the dielectric film as measured by $S'$.
 $R'$ and $T'$ are co-efficients of reflection and transmission measured in $S'$ respectively, where $R'= r'^{2}= \left(r'_i\right)^2$ and $T'= 1-t't'_{i}$.\\

So, using the Fresnel equations used in $S'$ gives

		 \begin{center}
		 \begin{equation}
		 \label{eq17}
		  r'\ =\frac{{E'_r}_\perp}{{E'}_\perp}=\frac{\cos i'-n'\cos r'}{\cos i'+n'\cos r'}
		  \end{equation}\end{center}
and
\begin{center}
		  \begin{equation}
		  \label{eq18}
		  t'=\frac{{E_t'}_\perp}{{E'}_\perp}=\frac{2\cos i'}{\cos i'+n'\cos r'}  . 
		 \end{equation}
		 \end{center}
Here, $n'$ is the refractive index of the dielectric film that corresponds to the Doppler shifted frequency $ \nu'$, which is given by Eq. (\ref{eq1}).\\
Using Stoke's relations in frame $S'$ \cite{Hecht}, we have $1-t't'_{i}=r'{^2}={r'_{i}}^2$ or $ R'+T'=1  . $

\begin{flushleft}

 Now, in the situation described so far, the relation $ R'+T'=1 $ does hold. As the direction of the velocity of the dielectric film is parallel to the air interface, the total energy of the incident wave is carried away by the reflected and transmitted waves. Otherwise this relation will not hold \cite{Energyjoss}.
    
Now, using inverse Lorentz transformation of $\textbf{E} $ field, the amplitude of $\textbf{E} $ vector of the first reflected beam in $S$ is 
\end{flushleft}

\begin{equation}
\label{eq19}
 {E_r}_\perp=\gamma\left(r' E'_\perp+\beta cB'_{r} \sin i'\right).
  \end{equation} 
Where $B'_{r}$ is the amplitude of the associated \boldsymbol{$B$} field of the reflected ray, as measured in $S$. The plus sign is due to the fact that the cross product of velocity and the vertical component of \boldsymbol{$B'_{r}$} is in the same direction as \boldsymbol{$ r'{E'}_\perp $}. \\
Then, as $ r'{E'}_\perp=B'_{r}c$ and using previously derived relation
   $ E'_\perp =\gamma E_\perp\left(1-\beta \sin i\right) $, it can be written that
   \begin{equation}
   \label{20}
       {E_r}_\perp=E_\perp\gamma^2r' \left(1-\beta\sin i\right)\left(1+\beta \sin \ i'\right) .
   \end{equation}
 So the effective reflectivity is $\gamma^2r' \left(1-\beta\sin i\right)\left(1+\beta \sin \ i'\right) $. In the non-moving situation,  $ \beta=0 $ and $ \gamma=1$, therefore the expression of the effective reflectivity reduces to a familiar form of
\begin{large} $\frac{\cos i- n_{0}\cos r}{\cos i+ n_{0}\cos r}$\end{large}, where $ n_{0}$ is the refractive index of the film in the non-moving case, corresponding to frequency $ \nu_{0} $ as measured by $S$.
 \\

To find the next $ \textbf{E} $ fields, we can just replace $ r' $ by $ t' t'_{i} r'_{i}, r'_{i}r'_{i}r'_{i}t' t'_{i} $ and so on. We define $\gamma^2\left(1-\beta\sin i\right)\left(1+\beta\sin i'\right)= u $.\\
If the film is sufficiently long, and if the incidence of light is not too oblique, then there will be a large number of waves. We can assume them to be infinite in number.\\

Therefore, adding up all the contributions of the $  \textbf{E} $ vectors in $S$ gives 
 \begin{equation}
 \label{eq21}
 \sum E_{r\perp}= ur' E_{\perp}\left(1-\frac{\ T'\ e^{i\Delta k_r}}{1-R'e^{i\Delta k_r}}\right).
 \end{equation}

Here $k_r$ is the angular wave number of the reflected wave as measured in $S$ given by  $2\pi / \lambda_r$ with $\lambda_r$ being given by Eq. (\ref{eq11}). $ \Delta $ is the path difference between two neighboring rays, it is given by Eq. (\ref{eq8}).\\

Using $ R'+T'=1 $, it follows that,
 \begin{equation} 
 \label{eq22}
\sum E_{r\perp}= uE_{\perp}r'\ \left(\frac{1-e^{i\Delta k_r}}{1-R'e^{i\Delta k_r}}\right)   .
\end{equation}
We now multiply by complex conjugates to find the total intensity pattern; after some basic algebra, it quickly follows that, the intensity of the reflected beam equals
   \begin{equation}
   \label{eq23}
   I_r =I_0u^2 \frac{4R'\ \sin ^2\left(\frac{\Psi}{2}\right)}{\left(1-R'\right)^2+4R'\ \sin ^2\left(\frac{\Psi}{2}\right)}   .
 \end{equation}
Here, $ \Psi$=\begin{large}$\frac{2\pi}{\lambda_r}$ \end{large}$\Delta $, the phase difference between two neighboring rays. $\lambda_r$ and  $ \Delta $ are given by Eq. (\ref{eq10}) and (\ref{eq8}).\\
  Now, writing $F_n=$\begin{large}$\frac{4R'}{\left(1-R'\right)^2}$\end{large}, it follows that
   \begin{center}
    \begin{equation}
    \label{eq24}
    I_r =I_0u^2\ \frac{F_n\sin ^2\left(\frac{\Psi}{2}\right)}{1+F_n\sin ^2\left(\frac{\Psi}{2}\right)}  .
    \end{equation}
   \end{center}
From energy conservation, it can be written $ I_0=I_r+I_t $  .\\
So, 
\begin{equation}
\label{eq25}
 	   I_t=I_0\frac{1+F_n\left(1-u^2\right)\sin ^2\left(\frac{\Psi}{2}\right)}{1+F_n\sin ^2\left(\frac{\Psi}{2}\right)} .
 	   \end{equation}
 	   \vspace{1cm}
In non-moving case, $ u^{2}=1 $. Then, $  I_r$ =$I_0$\begin{large}$\ \frac{F\sin ^2\left(\frac{\delta}{2}\right)}{1+F\sin ^2\left(\frac{\delta}{2}\right)} $ \end{large} and  $I_t$= $I_0$ \begin{Large}$
\frac{1}{1+F\sin ^2\left(\frac{\delta}{2}\right)}$
\end{Large} \cite{Hecht}.\\
Where $\delta$ =\begin{large}$\frac{4\pi d}{\lambda_{0}}$\end{large}$\sqrt{{n_0{^2}-sin^2{i}}}  $,
$F=$\begin{Large}$\frac{4R}{\left(1-R\right)^2}$\end{Large} and $R$ is the co-efficient of reflection in the non-moving situation, $n_0$ is the refractive index which corresponds to the  frequency $ \nu_0 $ as measured in $S$. So there is a clear change in the reflection and transmission pattern in the moving situation.

\section{Fabry Perot Spectroscopy with a Relativistic Medium}
\label{sec3}
 In the previous two sections, we have explored the effects of relativistic motion of optical medium on the conditions of constructive and destructive interference and reflection and transmission pattern. We have seen that the new conditions differ a lot from the non-moving situations. Finally, we will now turn our attention to the final inquiry.\\
The widely used FP spectroscope is built on the ideas of multiple beam interference. It is known for its extremely high resolving power compared to others (Michaelson-Morley interferometers, prism, N-slit grating etc). Now we will investigate how the movement of the medium (situated inside the FP etalon) affects the performance of the spectroscope.\\
Consider the most basic setup of the FP etalon, that is a slab of a particular dispersive medium is sandwiched by two mirrors with high co-efficient of reflection $ R $ \cite{fpeq}, as measured in lab frame $S$. Now the medium is made to move with speed $ v $ parallel to the walls of the mirrors with respect to $S$. Let us consider near normal incidence, hence incident angle $ i=0 $.\\
Now, we will investigate the propagation of light within the etalon. At first, light falls normally on the first mirror, the Huygens sources on the mirror act as primary sources, then the moving medium will act as the secondary Huygens sources according to the Huygen- Fresnel principle. Because of normal incidence, the Huygens sources on the mirror are all in phase in both lab frame $S$ and moving frame $S'$. Now, some of the incident light will be reflected and some will be transmitted through the mirror.\\  
Even though the two mirrors completely envelop the dielectric slab, there actually still remains an extremely little vacuum gap between the Huygens sources of the mirror and the moving medium. Hence, even though the medium sees a light wave tilted to the vertical by an angle $ \tan ^{-1} ({\gamma \beta})$ (this is found by by inserting $ i=0 $ in Eq. (\ref{eq2})) impinging on them, as the gap between the mirror and the medium is infinitesimally small, the disturbance that is forwarded on by the Huygens sources on the mirror is almost immediately received by the medium. So, in $S'$ the oblique incidence will not be factor due to the extreme proximity between the mirror and the medium. All the secondary Huygens sources of the medium would all be in phase, as they all received the disturbance simultaneously.\\
Thus in $S'$, the light travels down to the second mirror as shown in Fig. 5. Due to time dilation, the clock of the mirror runs slower, so the moving medium receives a light of frequency $ \nu_0/\gamma $ in their frame.\\
\begin{figure}[h]
\includegraphics[scale=.28]{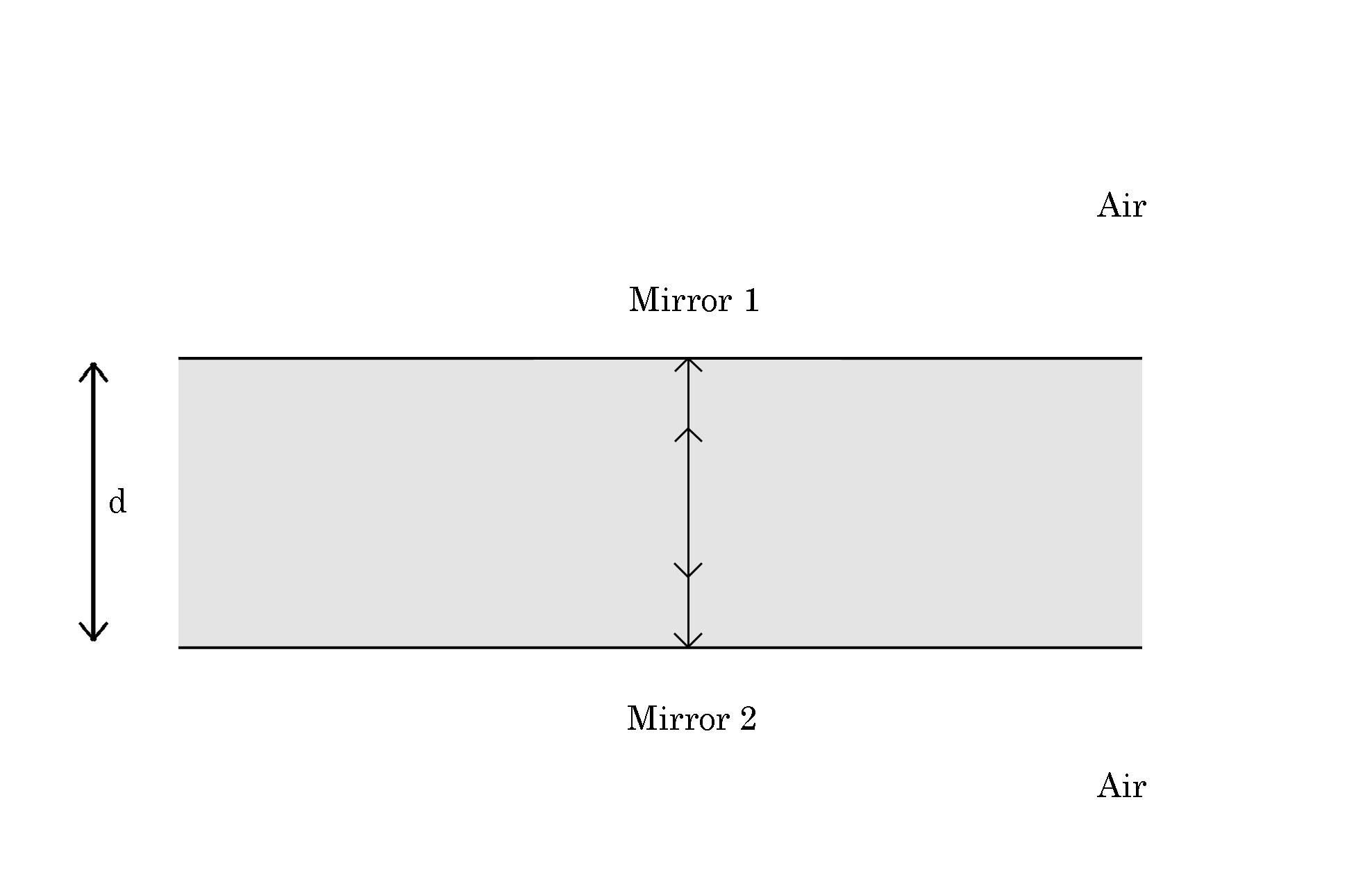}
\label{fig6}
\centering
\caption{The trajectory of light within the FP etalon, as seen by $S'$. In that frame, light rays of frequency $ \nu_0/\gamma $ travel vertically up and down between the two mirrors without any deviation.}
\end{figure}

When the transmitted light reaches the bottom, the Huygens sources of the moving medium then forward it onto the extremely small vacuum gap that exists prior to the second mirror. Within the gap, the direction of the forwarded disturbance is vertical in the moving frame $S'$. So, the second mirror (whose frame is analogous to the lab frame $S$) observes a tilted light wave by an angle $ \tan ^{-1} ({\gamma \beta})$ with the old frequency $ \nu_0 $. This claim can be substantiated by writing a 4 vector $\textbf{K}=(\vec{K},i\frac{\omega}{c}) $ and doing basic Lorentz transformation of the last component.\\
So, within that vacuum gap, in the frame of the moving medium $S'$,
\begin{equation}
\label{eq26}
\textbf{K}=( 0, -\omega'/ c, 0, i\omega' /c ),
\end{equation}

where, $ \omega'=\ 2\pi\nu_0/\gamma $. And as for the mirror, $\textbf{K}=( K_x, K_y, K_z, i  \omega_{mirror}/c)$.\\
For the 4th component of the 4 vector, the transformation is as follows
\begin{equation*}
i \frac{\omega_{mirror}}{c} = \gamma (\frac{i \omega'}{c}+i \beta . 0)= i \frac{\omega'\gamma}{c}   .
\end{equation*}
Dropping \begin{large}$ \frac{i}{c} $\end{large} from both sides and recalling $ \omega'=2 \pi \nu_0/\gamma $, it quickly follows that \begin{large}
 $\omega_{mirror}= \omega_{0}$, 
\end{large}
Hence, \begin{large}
$\nu_{mirror}= \nu_{0} $.
\end{large} 
\\

Again considering the reasons stated previously, all the Huygens sources of the second mirror will normally transmit and reflect the light to the air and the medium respectively. The moving medium will once again see a oblique light wave with frequency $ \nu_0/\gamma $. And in it's own frame, it will carry the disturbance back to the first mirror normally. This process will continue to repeat itself.\\
Now, in $S'$ the time taken by a light ray to complete a round trip back to the first mirror is simply, $ \Delta t'= 2n'd/c $ and as it comes back to where it started from, $ \Delta x'=0 $. And, as always, this $ n' $ corresponds to the frequency observed in $S'$, which is $ \nu_0/\gamma $.\\
So in $S$, using $ \Delta x'=0 $ and $ \Delta t'= 2n'd/c $, it follows that $$ \Delta t= \gamma(\Delta t'+\frac{\beta}{c} \Delta x')= \frac{2dn'\gamma}{c} $$ and
$$ \Delta x= (\Delta x'+\beta c \Delta t')\gamma= 2n'd\beta\gamma. $$\\\\
The light rays that emerge out of the etalon are normal to the mirrors. So, the effective path difference between two emerging light ray measured in $S$ is $ \Delta=c\Delta t=2n'\gamma d  $. Hence the phase difference between them is 
\begin{equation}
\label{eq27}
\Psi_{S}=  \frac{4\pi }{\lambda_0}n'\gamma d.
\end{equation} 
If the transmitted rays are collected by a lens, then the condition for constructive interference is
\begin{equation}
\label{eq28}
  \Psi_{S}= 2m \pi. 
\end{equation}

Where $m= 1, 2, 3 ...$ etc. Similarly, the condition for destructive interference is
\begin{equation}
\label{eq29}
  \Psi_{S}= (2m+1) \pi \ 
\end{equation}
where $m= 0, 1, 2, 3 ...$ etc.
In the non moving case, $ \Psi_{S}$ = \begin{large}$\frac{4\pi n_0 d}{\lambda_0} $ \end{large}, here $ n_0 $ is the refractive index for frequency $ \nu_0 $.\\
Therefore is a clear change of phase in these two situations. It has changed by a factor of $ \gamma n'/n_0 $. Hence, a fringe shift will occur.\\

The fringe shift will be equal to 
\begin{equation}
\label{eq30}
 \Gamma = \frac{2d\left(\gamma n'-n_0\right)}{\lambda_0} .
\end{equation}

This property may be used to measure the speed of the medium with high accuracy.
\\The movement of the dispersive medium in the etalon also affects the resolving power. To calculate the new resolving power, we will use the modified Rayleigh criterion, that is two wavelength components will be just resolved if the intensity dip at the center of the overlapped pattern is $ 8/\pi ^2 $ of the intensity of total maximum \cite{Rayleigh}.\\
Now, a change in $ \Delta\lambda $ would correspond to a change in $ \Psi_{S} $. The change is
\begin{equation}
\label{eq31}
 \Delta\Psi_{S} = \frac{4\pi n'\gamma d}{\lambda_0^{2}}\Delta \lambda_0.
 \end{equation}
Then some basic rigorous calculation leads to the relation \cite{Hecht}.
\begin{equation}
\label{eq32}
\Delta\Psi_{S} = \frac{4.147}{\sqrt{F}}.
\end{equation}
Where $ F$ is the so called finesse co-efficient of the spectroscope, which is equal to \begin{Large}$ \frac{4R}{\left(1-R\right)^2} $\end{Large}.\\
Using Eq. (\ref{eq31}) and (\ref{eq32}), the new resolving power becomes
 \begin{equation}
 \label{eq33}
\frac{\lambda_0}{\Delta \lambda_0}=3.03\frac{n'\gamma d\sqrt{F}}{\lambda_0}   .
 \end{equation}
Previously in the non moving situation it was  $3.03$ \begin{Large}$\frac{n_0 d\sqrt{F}}{\lambda_0}$\end{Large}. Hence, it has changed by a factor of $ \gamma n'/n_0 $. \\
From Eq. (\ref{eq31}), it's clear that the resolving power, which is defined as \begin{large}
$\frac{\lambda_0}{\Delta \lambda_0}$
\end{large}, is proportional to the phase difference of two neighboring light ray $\Psi_{S}$. So, as $\Psi_{S}$ changed by a factor of $ \gamma n'/n_0 $, the resolving power changed by the same factor.
\section*{Conclusion}
We have conducted a systematic inspection on the effects of movement of optical medium on optical interferometry. We have answered all the questions set forth in Introduction.\\ In section \ref{sec1}, a new modified condition for constructive and destructive interference in the case of a moving thin film has been derived. Next, in section \ref{sec2}, the expression of reflection and transmission pattern of light from a moving dielectric plane parallel film has been calculated. All of the new modified equations differ significantly from the previous non-moving situations and also reduce back to them if $\beta=0$.\\
In section \ref{sec3}, the resolving power of a FP spectroscope with a moving medium in its etalon has been derived, which differs by a factor $ \gamma n'/n_0 $ for normal incidence with respect to the non-moving case. Prior to this, the new condition of constructive and destructive interference for FP interferometer was also derived and a fringe shift was predicted from the pattern corresponding to the non-moving situation. All of these results can be momentous if $\beta$ is really high.\\
The main intention of the paper is to inspire an undergraduate student coalesce his/her concept of physics learned throughout different courses. A student must learn to ask and propose questions of research interest, and be able to clearly interpret phenomenons and build theoretical models, which usually requires a fusion of concepts of physics and mathematics. The exercise of these practices should start from early years of college, or even high school. This paper may serve as an introductory example to such a practice.\\\\

\section*{Acknowledgments}
The author would like to thank Emroz Khan for his valuable discussion and comments on this topic. And also Anuj S. Apte and Jason Kristiano for checking the calculations, and Fahim Murshed for the figures.\\\\
\section*{References}
\bibliographystyle{iopart-num}
\bibliography{arXiv}
\end{document}